# SD-WAN Threat Landscape

**Sergey Gordeychik**
serg.gordey@gmail.com
Inception Institute of Artificial Intelligence
Abu-Dhabi, UAE

**Denis Kolegov**
d.n.kolegov@gmail.com
Tomsk State University
Tomsk, Russia

## ABSTRACT

Software Defined Wide Area Network (SD-WAN or SDWAN) is a modern conception and an attractive trend in network technologies. SD-WAN is defined as a specific application of software-defined networking (SDN) to WAN connections. There is growing recognition that SDN and SD-WAN technologies not only expand features, but also expose new vulnerabilities. Unfortunately, at the present time, most vendors say that SD-WAN are perfectly safe, hardened, and fully protected.

The goal of this paper is to understand SD-WAN threats using practical approach. We describe basic SD-WAN features and components, investigate an attack surface, explore various vendor features and their security, explain threats and vulnerabilities found in SD-WAN products. We also extend existing SDN threat models by describing new potential threats and attack vectors, provide examples, and consider high-level approaches for their mitigations. The provided results may be used by SD-WAN developers as a part of Secure Software Development Life Cycle (SSDLC), security researchers for penetration testing and vulnerability assessment, system integrators for secure design of SD-WAN solutions, and finally customers for secure deployment operations and configurations of SD-WAN enabled network.

The main idea of this work is that SD-WAN threat model involves all traditional network and SDN threats, as well as new product-specific threats, appended by vendors which reinvent or introduce proprietary technologies immature from a security perspective.

# TABLE OF CONTENTS

# 1 INTRODUCTION

The software defined wide-area network is a technology based on SDN approach applied to WAN connections in Enterprises. According to Gartner's predictions, more than 50% of routers will be replaced with SD-WAN solutions by 2020 but community knowledge on SD-WAN threats and attacks is limited.

New systems and components deployed in networks to carry out SD-WAN functions may become a target of attack and compromise network security of all organisations. By definition, an SD-WAN system forms network perimeter and connects Internet, WAN, extranet, and branches that makes it attractive targets for attackers. SD-WAN can have firewalls, DPI, VPN, malware detection and other traditional security features on board which is crucial from a cybersecurity point of view.

Previous works [1 - 6] focus on general SDN technology or Virtual Networks Functions (VNF) threats and attack surface. The goal of this paper is to highlight practical aspects of SD-WAN threat modeling applied to real-world well-known products.

We explored the most widespread and popular SD-WAN products employed in enterprise and corporate networks. Table 1 lists SD-WAN vendors and products we examined in our research. Versa, Citrix, and Cisco products were deployed locally. The other products were deployed on AWS infrastructure.

Table 1. Explored SD-WAN Vendors and Products

| Vendor | Products |
| --- | --- |
| Versa Networks | Analytics<br>FlexVNF<br>Director |
| Citrix | NetScaler SD-WAN Center<br>NetScaler SD-WAN (SD-WAN) |
| Talari | Cloud Appliance |
| Viprinet | Virtual VPN Hub |
| Silver Peak | Unity Orchestrator<br>Unity EdgeConnect |
| Riverbed | SteelConnect<br>SteelHead |
| Cisco (Viptela) | vManage<br>vBond<br>vSmart<br>vEdge |

## Approach Overview

Our approach is based on the following international standards and best practices:
1. Open Web Application Security Project (OWASP) Testing Guide.
2. Penetration Testing Execution Standard (PTES).
3. NIST Special Publications 800-115 Technical Guide to Information Security Testing and Assessment.
4. Center for Internet Security (CIS) standards.

We basically reviewed product documents (data sheets, configuration and operation guides, slides, etc.) and reference specifications manually and tested real systems in simple scenarios.

The steps we employed to identify threats, security weaknesses, and potential vulnerabilities for each SD-WAN product are as follows:
1. The product documentation and specification reviewing.
2. The product deployment. As a rule, we deployed a network consisting of two branch offices connected by several WAN links. That testbed enabled us to send network packets over the network.
3. Software decomposition and inventory.
4. Entry points, interfaces and protocols enumeration.
5. Information leakage analysis and fingerprinting.
6. Operating system hardening measure and mechanism analysis.
7. Security feature investigation and analysis.
8. Multi-tenancy analysis.
9. Security assessment within penetration testing approach for some components and functional blocks.
10. Threat modeling, documenting, and responsible disclosure.

It should be noted that we did not use the above steps for all the SD-WAN products from the scope due to limited resources. In some cases we confined ourselves to management plane security analysis.

## Contributions

The contributions of the paper include the following:
1. A taxonomic threat modelling approach for SD-WAN systems.
2. A practical SD-WAN system threat model extending the previous SDN threat models within practical aspects.

This approach allowed us to uncover multiple design flaws, security issues, weaknesses, and zero-day vulnerabilities in the SD-WAN products.

## Related Work

Yoon et al. [1] conduct a comprehensive survey of possible methods for abuse or exploitation of an SDN stack based on OpenFlow, create a basic SDN attack surface, provide a generalized classification of attack vectors for abuse or direct attack on SDN. However, they consider only control and data plane attack surfaces and do not analyse management and orchestration planes at all.

Several research papers provide theoretical reports related to SDN threats. Hizver [2] tries to systematically identify SDN threats. He enumerates threat sources, vulnerability sources, threat actions and then describes the integrated SDN threat model. At the same time that work is theoretical and does not provide any real examples of vulnerabilities or attacks on SDN. Moreover, the report provides common attacks that are specific not for SDN, but for general computer systems. The following examples were described: disruption of service using flooding attacks, unauthorized access using password brute-forcing or password-guessing attacks, unauthorized access using remote application exploitation attacks, etc. The same approach is applied in [3]. They focus on container-based virtualization approach and consider Container-as-a-Service platform for SDN within NVFI architecture. The report also analyses security threats and provides mitigation strategies to secure VNFs.

Shankar [4] describes critical security threats that exist in the NFVI and proposes best common security practices to be protected against them. The following high-level methods are considered: VNF image signing, kernel hardening, remote attestation, isolation, secure booting, etc. It should be pointed out that we have not found an SD-WAN product implementing those proposed methods.

Rajendran [5] focuses on security analysis of Nuage VNS solution. He describes the attack surface and security weaknesses that were found in Nuage VNS product. He also discloses vulnerabilities in the CPE zero-touch provisioning implementation and demonstrates some man-in-the-middle attacks exploiting these vulnerabilities.

The Verizon SDN-NFV Reference Architecture [6] provides 8 layers of threat vectors associated with securely delivering a service in a network based on SDN, NFV and Virtualization. That document provides common requirements and basic recommendations regarding secure communication, authentication, audit, etc. The Verizon Reference Architecture contains common and highly useful recommendations that are not used and followed by vendors, as we will see below.

## Organization of the Article

The remainder of this article is organized as follows. Section 2 considers SD-WAN architecture, design principles, and the main components. We also discuss common approaches to software implementation and open source reuse. Section 3 addresses threat enumeration for SD-WAN. We describe most frequent threats and provide examples of found vulnerabilities and possible attacks. The main results are given in Section 4.

## Terms and Definitions

For the purposes of the article, we will use the main terms and definitions of documents [6 - 8].

# 2   SD-WAN TECHNOLOGY OVERVIEW

**SD-WAN Components**

SD-WAN enables new implementation of the planes and its functions on the SDN-NFV infrastructure specific to WAN which provides additional features for customers, such as:
- Multi-tenancy (VRF and routing)
- Zero-touch provisioning
- Overlay and dynamic tunneling VPN
- WAN optimization
- Intelligent path selection and route learning
- Automatic bandwidth detection
- Service chaining
- Security services such as VPN, NGFW, WAF, URL categorization and filtering, DPI/IDPS, SSL/TLS inspection

To fully understand SD-WAN threats an understanding of overall architecture, component interconnection patterns, typical control and data flows is required. Any of these can contain vulnerabilities and be exploited by attackers to compromise an enterprise network. Figure 1 illustrates a general SD-WAN architecture and its building blocks.

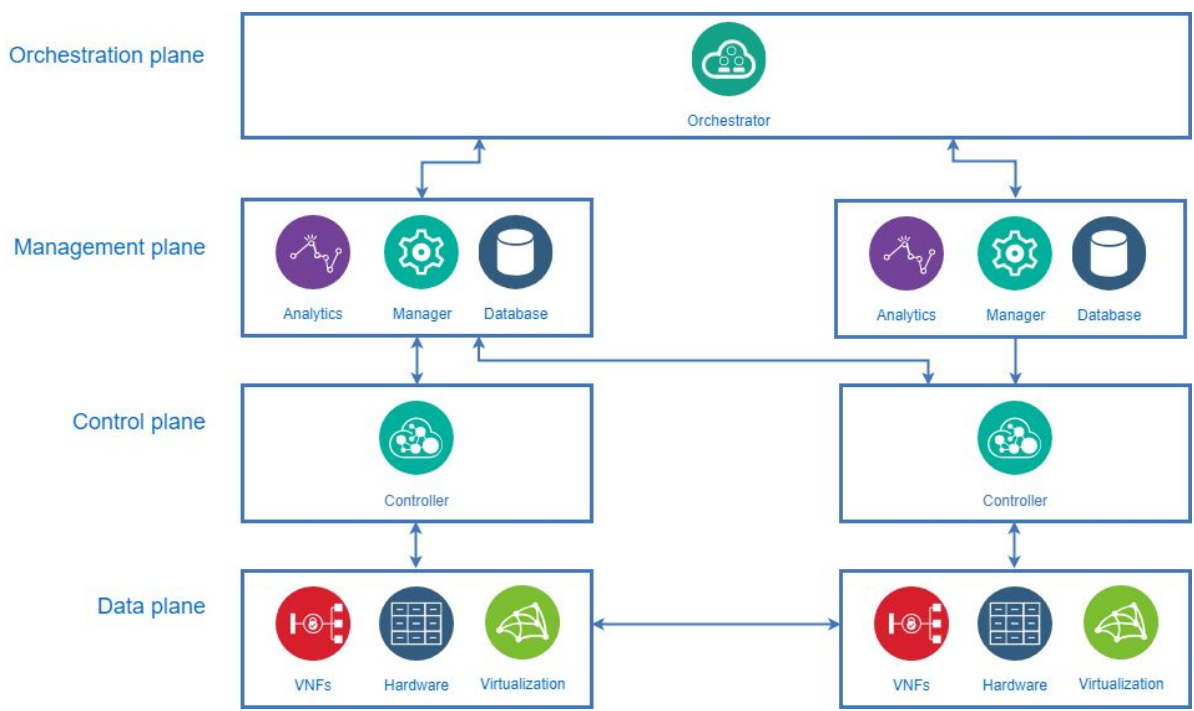

Figure 1. SD-WAN basic logical architecture and flow

Based on the SDN-NFV reference architectures [6 - 8], the following main functional blocks were identified in SD-WAN:
- Network Function (NF): functional block within a network infrastructure that has well-defined external interfaces and well-defined functional behavior.
- Network Service (NS): composition of Network Functions defined by its functional and behavioural specification.
- Virtualized Network Function(VNF): implementation of an NF that can be deployed on a Network Function Virtualisation Infrastructure (NFVI). The following are examples of VNFs: DPI, IDPS, WAF, LB, NAT, PROXY, and VPN.
- NFV Infrastructure (NFVI): hardware and software in which VNFs are deployed.
- Network Controller (CTRL): functional block that centralizes some or all of the control and management functionality of a network domain and provides an abstract view of its domain to other functional blocks via well-defined interfaces.
- Network Functions Virtualisation Orchestrator (NFVO): functional block responsible for the management of the NS life cycle, VNF lifecycle and NFV infrastructure resources.
- Network Functions Virtualisation Infrastructure Node (NFVI-Node): physical device(s) deployed and managed as a single entity, providing the NFVI Functions required to support the execution environment for VNFs.
- Network Service: composition of Network Function(s) and/or Network Service(s), defined by its functional and behavioural specification.
- VNF Manager (VNFM): functional block that is responsible for the lifecycle management of VNF.
- Network Element (NE): group of data plane resources that is managed as a single entity. Network element, router and edge router terms are used interchangeably.
- Management interface (MGI): functional block that is responsible for accessing and governing a set of resources and functions.
- Multi-tenancy: feature where physical, virtual or service resources are allocated in such a way that multiple tenants and their computations and data are isolated from and inaccessible by each another
- Tenant: one or more SDN-NFV service users sharing access to a set of physical, virtual or service resources

SD-WAN system implementation is not always based on a hierarchical model from a network topology point of view (Figure 2). For example, controller and orchestrator can be deployed in the same IP network or broadcast domain. This can allow an attacker to perform vertical access control attacks and access management interfaces and functions.

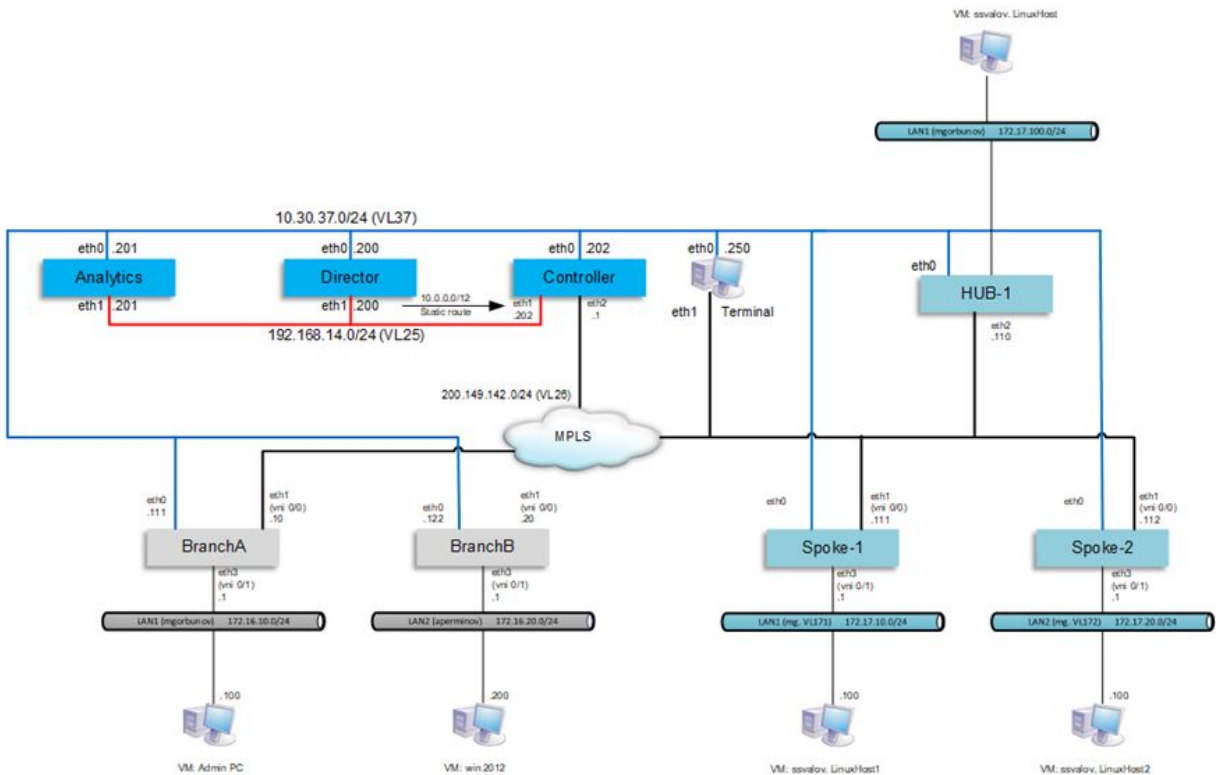

Figure 2. SD-WAN testbed network topology example

Different SD-WAN components, in turn, can have their own internal structure. For instance, a VNF may consist of one or multiple components, called VNFC [7]. A VNFC, in this case, is a software entity deployed as a virtual machine or container. At the same time, a VNF is managed as a single element even if it is implemented by a set of VNFC. Having said this, a separated internal network can be employed inside an SD-WAN component (Figure 3). This network provides VNFC connectivity and can process East-West traffic using sophisticated load balancing or failover algorithms. Those and other design statements should be addressed within threat modelling.

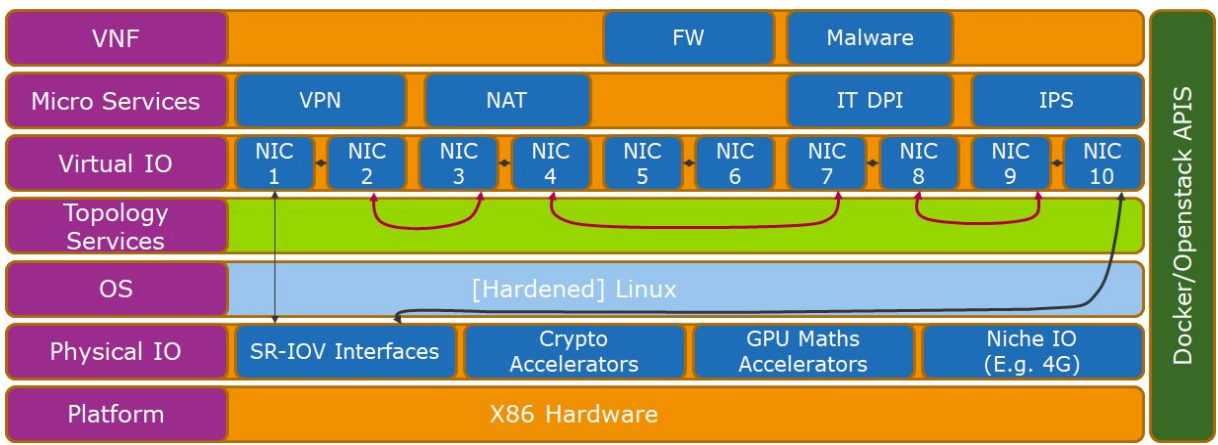

Figure 3. The example of SD-WAN Network Element with VNF support

# SD-WAN Component Implementation

From a software engineering perspective, SD-WAN components, such as Orchestrator, Controller or Network Element can be built on different open-source and proprietary software. We enumerate the main software components and their implementation, we have seen in the analyzed SD-WAN products, in Table 2.

Table 2. SD-WAN software component examples

| Component | Software |
| --- | --- |
| | |
| Virtualization | |
| Virtualization | OpenStack, KVM, Docker |
| Network services | |
| Packet processing | DPDK |
| Proxy | HAproxy |
| Routing | Bird, Quagga |
| Service discovery | Avahi |
| Northbound protocols | NETCONF, REST, XMPP |
| Southbound protocols | OpenFlow, OMP, MP-BGP, BGP, custom proprietary protocols |
| Data plane | VXLAN, GRE, MPLS, AH/ESP, TLS, DTLS |
| Management services | |
| SQL database | MySQL, MariaDB, PostgreSQL |
| NoSQL database | ElasticSearch, Cassandra, OrientDB, Apache Solr |
| Message broker | Apache Kafka |
| Web server | Apache, Nginx |
| Web application server | Apache Tomcat, Karaf, WildFly, Node.js |
| Web framework | Django, CakePHP |
| Cache | Redis |
| Resource monitoring | Munin |
| Remote management | AjaxTerm, OpenSSH, Shell In A Box |
| Front-end | Angular, ExtJS |
| Security Services | |
| Encryption | strongSwan |
| Firewall | netfilter/iptables |
| Web application firewall | Modsecurity, OWASP CRS rules |
| IDPS | Snort, suricata. ET Pro rules |
| DPI | Qosmos ixEngine |
| Filtering and classification | MaxMind, BrightCloud, Cyren services |
| Anti-virus | Cyren engine |
| DNS security | Farsight database |
| System Software | |
| Operating system | Ubuntu, Debian, custom distributive |

# 3 UNDERSTANDING SD-WAN THREATS

This section considers security threats related to the main SD-WAN functional blocks and components.

**Outdated Software**

In general, operating systems used in SD-WAN nodes (edge router, controller, orchestrator, etc.) are built on general-purpose GNU/Linux distributions like Ubuntu, Debian, CentOS as well as other network software products. In this regard, one of the most important question is existence of zero-day and known vulnerabilities in the kernel of the operating system and in other its software modules. It was shown [9] that many of the SD-WAN products leak information related software versions (e.g., operating system, kernel, build versions). For instance, in February 2018, Shodan search engine detected Silver Peak appliances with the operating system based on unsupported and outdated Linux 2.6.38, released in March 2011 (see Figure 4).

Figure 4. Linux kernel version leakage for Silver Peak products.

Those vulnerabilities, in general, provide local facilities and can be employed by attackers to escalate privileges in the operating system.

**IPMI and Out of Band Management**

SD-WAN systems can be delivered as a hardware appliance using commercially available off-the-shelf (COTS) platforms. Modern server platforms support interfaces used by system administrators for out-of-band management of computer systems and monitoring of operations, such as Intelligent Platform Management Interface (IPMI). Main risks of using the IPMI and Baseboard Management Controllers (BMCs) are as follows [10]:

- Passwords for IPMI authentication are saved in clear text.
- Knowledge of an IPMI password gives you password for all computers in an IPMI managed group.
- Root access on an IPMI system grants complete control over hardware, software, firmware on the system.
- BMCs often run excess and older network services that may be vulnerable.
- IPMI access may also grant a remote console access to the system, resulting in access to the BIOS.
- There are few, if any, monitoring tools that are able to detect if the BMC is compromised.
- The certain network flows to and from BMC are not encrypted.
- Unclear documentation on how to sanitize IPMI passwords without destruction of the motherboard.

Compromised BMC can be used by threat actors to establish persistence that lasts beyond an HDD wipe and reinstalling operating system. Attackers can use BMC trojans for lateral movement from host-side networks to management networks which are intended to be isolated.

It is worth noting that update management mechanisms for IPMI components on OEM-platforms do not satisfy most of modern security requirements. For instance, Citrix CTX216642 update [11], published in September 2016, identified and fixed more than 30 security vulnerabilities in Supermicro IPMI Lights Out Management (LOM); these vulnerabilities were known from May 2013.

## Basic Network Functions

A NE performs packet processing functions and can be implemented as hardware, software, virtual, or cloud appliance. Cloud NE concept is used within Virtual Customer Premises Equipment (vCPE) or Cloud Customer Premises Equipment (cCPE) models in service delivering.

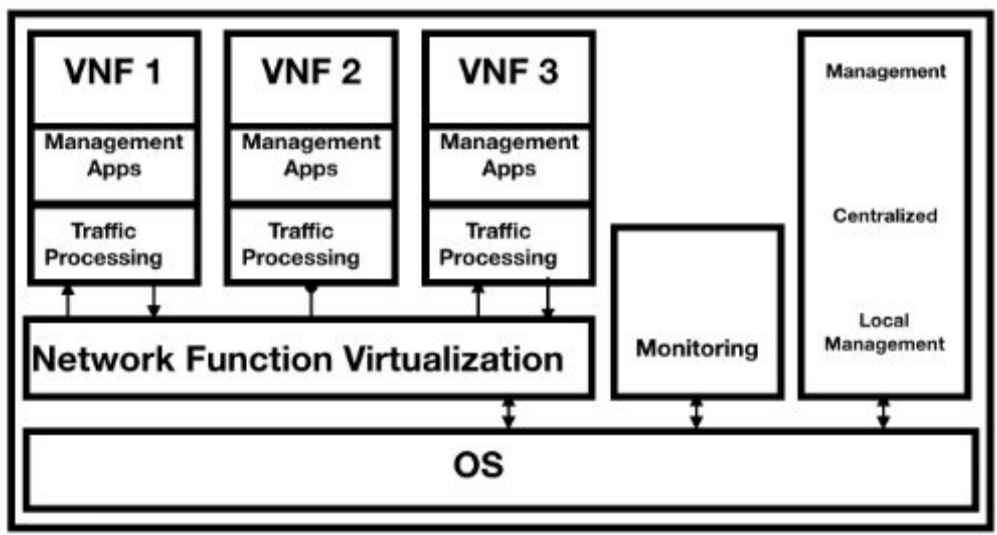

Figure 5. SD-WAN Network Element with VNF support

The following are the most important SD-WAN components used in East-West packet processing:
- *Packet Processor* based on software implementing kernel bypass techniques (DPDK, PF_RING, OpenOnload, etc. PDK).
- *Routing* based on traditional routing protocols (BGP, OSPF, RIP) and proprietary mechanisms like "SNMP Route Learning". Open-source software like Quagga and Bird are generally used in routing components.
- *Overlay* that involves running a logically separate network over traditional links using tunneling, encapsulation, and encryption methods. The overlay protocols also can be traditional (VXLAN, GRE, MPLS, etc. ) or proprietary.
- *VPN* that is used to provide secure transport within overlay protocols. The strongSwan's is the popular open-source implementation of IPsec used in most products. It is worth noting that strongSwan supports AH or ESP protocol, but

AH+ESP bundle is not supported. At the same time many SD-WAN use custom cryptographic protocols, that automatically arise the set of cryptography related threats.

The most important difference and peculiarity of SD-WAN packet processors from a security point of view is a necessity to consider threats targeting control plane via threats of management or orchestration plane. All explored SD-WAN products involved multiple web interfaces and generally an SQL database, that was shared between the services of different planes (management, control, forwarding, etc.). The same database generally stores information related to routing and forwarding, and, for instance, user interface, logging, statistic data. Network routes from the database are pushed to Routing Information Base and influence to Forwarding Information Base. For this reason, a vulnerability to SQL injection attack could allow an attacker to modify data in routing database, injects new routes and reroute network packets. It also means that in this case the attacker has given control under packet processor features implementing control flows from database to data plane features.

As a result, a packet processor implementation vulnerabilities could arise the following main threats:
- Using components with known vulnerabilities
- Sensitive information disclosure
- Arbitrary code execution
- Denial of service
- Network spoofing via ARP poisoning or ICMPv6 router advertisement attacks
- Traffic rerouting
- Poorly implemented data and control plane protocols

For example, according to [12], on August 14th, 2018, Arista released information about a denial of service vulnerability where a crafted IP fragment ordering or overlap can allow an attacker to consume more memory than defined in the Linux kernel settings.

Routing threats consist of target routing protocol (BGP, OSPF, etc.) well-known threats, such as BGP route leaking, route hijacking, and so on. It also includes reinvention routing mechanism product-specific threats. For instance, one of the SD-WAN products provided a SNMP route learning mechanism importing network routes from a remote router over SNMP v2c protocol. The security of this feature is based on SNMP community string strength and SNMP "requestID" randomness. So, it does not even implement a traditional integrity and authenticity mechanisms provided by all modern routing protocols. This mechanism must not be used within MitM attacker model: the SNMP community string mechanism cannot be used to protect routing information against modification; it enables protection against remote route spoofing attacks only.

SD-WAN VPN threats extend employed cryptographic protocol (IPsec, TLS, DTLS, etc.) threat in the following sense. First, they involve general implementation related threats like control flow integrity, denial of service. Second, they involve threats to cryptographic protocols. For example, IPsec protocol commits the following threats:
- Weak key and cryptographic parameter selection [13].

- Weak authentication, such as weak passwords for PSK or weak certificate or trust validation for RSA-based authentication.
- Weak IKE parameters selection [14, 15].
- Specific implementation threats, such as Bleichenbacher-style signature forgery which involves RSA padding attack (CVE-2018-15836) [15].

Third, in addition, SD-WAN products often employ custom cryptographic solutions as they employ custom network protocols. For example [16], Citrix NetScaler SD-WAN appliances provide IPsec and proprietary "Global Virtual WAN Network Encryption" mechanism. In this regard, the following additional threats arise:

- Poorly implemented cryptography mechanism
- Insufficient cryptography protection
- Customer identity disclosure
- Use of insecure cryptographic algorithm
- Peer impersonation

Forth, VPN orchestration features are becoming most important since most SD-WAN products implement overlay mechanism based on encrypted tunnels with automatic configuration and cryptographic parameter agreement. Those mechanisms are responsible for cryptographic parameter and algorithms selection between communication entities. Having said this, most realistic attacks are downgrade attacks based on spoofing or MitM attacks on data plane, and insecure configuration related threats on control plane.

It is recommended to apply the following security best practices to be protected against these threats:

- Use standard cryptographic and network protocols
- Update an operating system and third party components
- Employ confinement mechanisms like SElinux and AppArmor
- Use secure cryptographic parameters
- Employ secure cryptographic primitives
- Apply hardening-based mitigation methods
- Develop internal mechanisms detecting network attacks

### Network Functions Virtualization

Different approaches can be used to implement NFV mechanisms. These approaches include specialized VNF-aware applications within a base operating system and fully-featured hypervisor- or container-based virtualization mechanisms connected together in inline or passive modes in NFV environment. Given the fact, that a VNF processes transit packets, an intruder can attack forwarding, detection, or inspection engines on each VNF. For example, incorrect regular expressions in signature-based IDS (e.g., suricata) or WAF (e.g., modsecurity) can cause a vulnerability to Regular Expression Denial of Service attack. The example of this vulnerability is CVE-2017-15377 [17]. In the research, we identified a set of regular expression vulnerabilities in a SD-WAN product employing an "Emerging Threat" rule set.

Beside that, each VNF has a bunch of control mechanisms (e.g., configuration management, logging, etc.). The corresponding interfaces, as usual, are available to control plane via southbound interfaces and employed by controller or orchestrator nodes. We were able to uncover a bunch of logical vulnerabilities related to improper authorization that could allow us to gain control over a VNF operating system. At the time of this writing, the vulnerabilities are being fixed by corresponding vendor and we will disclose them in the next papers.

A compromised VNF, in turn, can be considered as the source of threats targeting other SD-WAN components. Since a VNF and its components are connected together using TCP/IP stack, insufficient or improper VNF isolation and confinement can allow an attacker to commit a wide range of network attacks against the control and data plane of another VNF. The examples include password guessing, SSRF, DoS, traffic manipulation, memory corruption attacks, and so forth.

As a rule, VNF is implemented on virtual machine or container. Accordingly, NFVI vulnerability may allow an attacker to escape from the confines of an affected virtual machine guest operating system and potentially obtain code-execution access to the host. Typical root cause examples involve vulnerabilities in virtualization mechanism implementation or configuration allowing an expanded access from a virtual machine guest to a host via different data exchange interfaces. For instance, the CVE-2015-3456 describes a vulnerability in unnecessary floppy disk controller in QEMU that allows local guest users to cause a denial of service (out-of-bounds write and guest crash) or possibly execute arbitrary code [18]. In bare metal server, one of the actual threat is exploitation of vulnerabilities to Meltdown [19] and Spectre [20] attacks.

We did not identify SD-WAN specific threats in NE and VNF management interface. These threats will be described in the corresponding section devoted to management plane threats. It should be noted, that vendors often use monitoring services (e.g., Munin) and apply insufficient security mechanisms to protect them. Generally, the services are bind to internal VNF network and accessible via the local interface (loopback interface) only. This approach is not sufficient in NFV architecture suggesting that several components with different trust level communicate with each other. As a result, threats regarding unauthorized access to a monitoring service and remote management (including operating system arbitrary command execution) can appear.

We identified the following main threats here:
- Denial of service via security function
- Unauthorized access to VNF resources
- Privilege escalation

Figure 5 illustrates privilege escalation attack vectors targeting control and management planes within an SD-WAN NE. The first step (1) is to get unauthorized access to a VNF by exploiting a vulnerability in Traffic Processing mechanism. Then the attacker can commit: an attack against an operating system (2), a horizontal privilege escalation attack against Traffic Processing mechanism of another VNF on the same NE(2'), or an vertical privilege escalation attack against management application on the same VNF (2''). Having got access, the attacker can use a horizontal privilege escalation attack against management application of another VNF (3), attack NE management (3'), or attack VNF management mechanisms (3'') from operating system layer. The last but not least here is a privilege escalation attack targeting of South - North communications exploiting control and management plane vulnerabilities of another SD-WAN nodes (4).

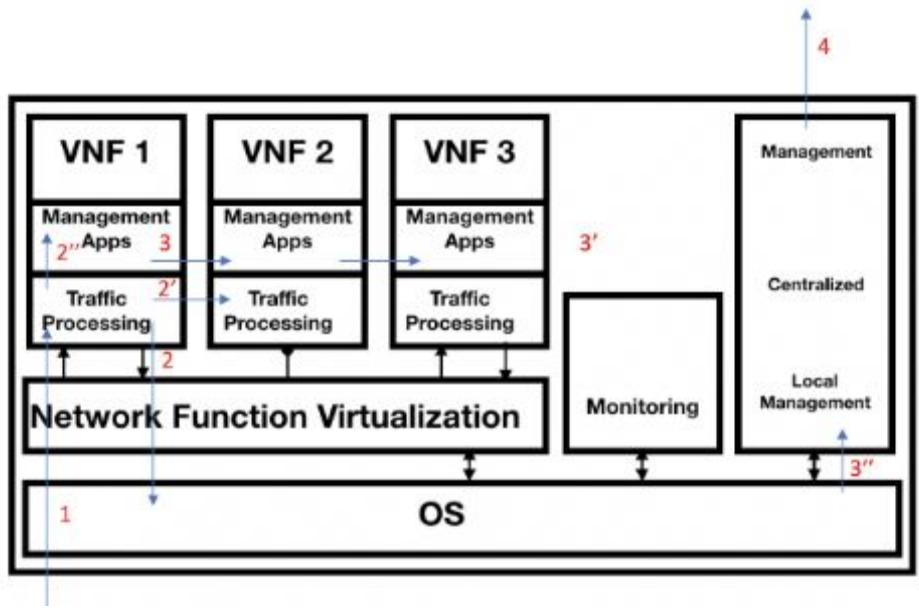

Figure 5. Privilege Escalation Attack Vectors Example

We think the last attack vector is the most exploitable and probable; most of VNF communicate with controller using rich API and network protocols. Our experience shows that even those features in SD-WAN always contain implementation or operation security issues and can be exploited.

**Management Interfaces**

Management plane traditionally involves network protocols (e.g., SSH, TLS, HTTP), management interfaces (as usual, command line interface (CLI) and Web user interface (Web UI)), and software (Web browsers, Web servers, SSH clients, etc.). Despite that all best practices and security guides recommend not to connect management interfaces to the Internet and use them in out-of-band or logically separated networks, SD-WAN management interfaces, by design, are used not only to communicate with system administrators, but also with other SD-WAN components like orchestrator and controller.

It was shown [9], that about 2 000 recognisable SD-WAN interfaces are connected on the Internet. Most of them are management interfaces. This result can be explained by moving SD-WAN to clouds and corresponding delivering SD-WAN via clouds such as AWS or Azure.

Some SD-WAN features such as Zero Touch Provisioning and Multi-tenant Management should have access to the Internet via management interface by design. We will discuss this matter further.

Management interfaces are divided into the following groups:
1. Local and remote console (CLI).
2. Human-machine web interfaces (Web UI).
3. Machine to machine (Northbound and southbound interfaces) API.

### Command Line Interface

CLI is used remotely and locally to interact with SD-WAN services. Most of them, as we have already known, employ GNU/Linux as a basic operating system thereby inheriting the general management mechanisms and tools like Telnet, SSH, command shells, and so on. In the most cases, restricted shell is used as a command shell. That shell limits user abilities and only allows to perform a specified subset of system commands or a subset of high-level commands. Another approach is to provide a special restricted account. These approaches are traditionally considered as a common best practice, but their implementations often contain the following weaknesses:
- Hardcoded accounts and passwords
- Weak default passwords
- A use of a special unlimited system user account that is able to access an operating system command shell directly
- Permissive sudo configuration
- Incorrect access control configuration
- Vulnerabilities in GNU/Linux components (local privilege escalation)
- Vulnerabilities in restricted shells

The examples are as follows:
- "_spsshell" command in Silver Peak allowing to access BASH shell directly [21]
- "CBVWSSH" user account in Citrix NetScaler SD-WAN used by design to access debugging facilities
- "admin" user account with predefined password in Cisco (Viptela) SD-WAN

A use of SSH and Telnet protocols for remote management causes well-known threats. At the same time, it was found that some SD-WAN products additionally employ web-based tools like "AjaxTerm" and "Shell in a Box". This extends an attack surface over web technologies and could allow an attacker to commit traditional Web attacks like DNS Rebinding, XSS, CSRF, Command Injection, and so on. Also, there are vendors (e.g., Viprinet) which use cleartext HTTP protocol without SSL/TLS.

### Web User Interface

All explored SD-WAN products employ web user interface (Web UI) as a main unified interface to govern all parts of the SD-WAN system (routers, controller, orchestrator, etc.) over its data, control, management, and orchestration planes.

It was found that existence of low-hanging critical vulnerabilities in SD-WAN related to Web UI is a widespread and systematic. We identified vulnerabilities to the following classic attacks:

- HTTP Slow DoS-attacks
- Password brute-forcing
- XSS (Reflected, stored)
- Command injection and RCE
- XXE and SSRF
- CSRF
- IDOR

For instance, the security bulletin [22] describes identified vulnerabilities in the Web UI of the Citrix SD-WAN physical appliances and virtual appliances. The vulnerabilities could allow an unauthenticated attacker having a network access to the management interface to compromise the entire SD-WAN network.

The other widespread security issues identified in Web UI are sensitive information leakage and vulnerabilities to SSRF. First, most of SD-WAN products provide logging facilities enabled by default. They often leak sensitive information such as user passwords and session tokens into world-readable files. These leakages can cause unauthorized access or privilege escalations. Second, since SD-WAN involves a set of interconnected network services and applications, from our point of view, a vulnerability to SSRF attack is one of the most critical vulnerabilities. SD-WAN services, as usual, belong to the same trust domain and an ability to send requests behalf on such services could allow an attacker to bypass authorization mechanisms and gain access to critical services like REST API (see the examples in the Network Functions Virtualization Threats section).

## Multi-tenancy

An SD-WAN system is called multi-tenant, if it has an ability to provide logical isolation of shared resources within NFV concept. This suggests an existence of common Web UI used by different tenants to access isolated SD-WAN instances. It turned out that those Web UI also have web security vulnerabilities specific to multi-tenant applications such as missing or improper access control. In this context, it is worth noting that an adversary having a legitimate access to the management interface can commit threats against a provider or other clients (tenants). For example, standard or weak passwords for different tenants or weak password recovery mechanism can lead to horizontal privilege escalation. In this regard, the adversary can examine implementation and its possible weaknesses using an own isolated session and then exploit found vulnerabilities against other clients.

During the research, the following main threats regarding multi-tenancy were identified:
- Unauthorized access to provider's data
- Unauthorized access to tenant's data
- Unauthorized access to tenant's VNF
- Unauthorized access to tenant's stored flow data (e.g., NetFlow, IPFIX)
- Denial of Service

The following threats are most actual regarding management plane:
- Unauthorized access to management interface on the Internet

- Enumeration and fingerprinting SD-WAN nodes on the Internet
- Exhaustive Denial of Service
- Privilege escalation
- Insufficient authentication
- Insufficient access control
- Local or remote privilege escalation within operating system access control

The following common recommendations can be made to protect management:
- Do not use default passwords
- Apply secure provisioning during an initial configuration of a node
- Network service confinement and isolation
- Input validation and output escaping for data crossing trust level boundary
- Systematically update own and third-party software components
- Correctly configure TLS protocol
- Apply secure software development lifecycle practices
- Employ embedded WAF mechanisms as defence in depth mechanism protecting Web UI
- Harden an operating system components

## Orchestration Services

### Service Interfaces

Northbound interface is used to interface with orchestration layer. As a rule, SD-WAN vendors prefer to employ REST API, XML-based, or custom protocols. Most used protocols are HTTP or Websocket.

Southbound interface is used by controllers to communicate with to communicate with routers. In general, SD-WAN employs Openflow, NETCONF, REST API, or custom protocols over TLS to program the network connectivity and GRE, VXLAN to provide overlay tunnels.

The main common threats which are most critical to our opinion and were identified in the research include the following:
- Unauthorized access to a dangerous method or function via exposed interface
- Unauthorized access to operating system resource via vulnerable functions or methods
- SD-WAN node (edge router, controller, orchestrator) spoofing
- Sensitive information exposure
- Eavesdropping

Let's consider the most critical vulnerability we managed to identify within the research in the context of interface threats. We found a use of hard-coded TLS certificate and the corresponding public-key cryptography key pair in a well-known SD-WAN product (at the time of writing this vulnerability has not been fixed yet). It turned out that all SD-WAN controllers and orchestrators of this vendor use the same pre-installed self-signed RSA certificate in a controller - orchestrator communication protocol. An attacker in passive MitM position could use the certificate and corresponding private key to perform eavesdropping and spoofing attacks against all SD-WAN nodes. Using this vulnerability the attacker can commit all above threats.

In another case, a vendor used Websocket-based protocol to implement a Northbound interface. The implementation were vulnerable to Cross-Site Websocket Hijacking attack, As a result, an authenticated remote attacker could perform arbitrary command on a edge router.

## Zero Touch Provisioning

Zero Touch Provisioning (ZTP) or Zero Time Deployment (ZTD) is a mechanism that allows nodes to be provisioned and configured automatically. For example, Citrix SD-WAN ZTD mechanism enables an SD-WAN appliance to communicate with Citrix ZTD Cloud Service that is publicly accessible from any point on the Internet using management interface [23].

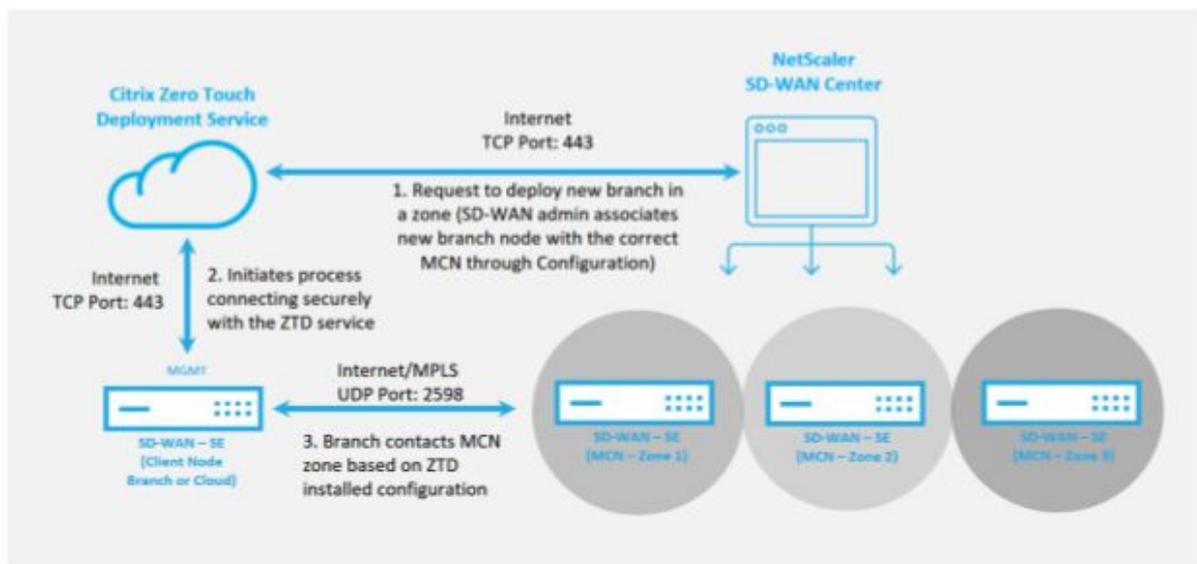

Figure 6. Citrix ZTD High-Level Architecture.

It is obvious, that in that case authentication and authorization mechanisms of ZTP service are critical elements of SD-WAN security. For example, CVE-2018-0434 [24] describes a critical vulnerability in Cisco SD-WAN Zero Touch Provisioning feature that could allow an unauthenticated, remote attacker to gain unauthorized access to sensitive data by using an invalid certificate. An attacker could exploit this vulnerability by supplying a crafted certificate to an affected device. A successful exploit could allow the attacker in MitM position to decrypt confidential information between SD-WAN device and ZTP service.

Often, ZTP services are implemented as Web services and contain typical vulnerabilities to attacks such as SQL Injection, OS Command Injection, SSRF, and so on.

At the same time, by design, ZTP server must accept requests from unidentified and unauthenticated devices on the Internet. This fact increases attack surface and attacker capabilities. As an example, let's consider a vulnerability CVE-2018-0347 [25] that could allow an authenticated, local attacker to inject arbitrary commands that are executed with root privileges. A successful exploit could allow an attacker to execute commands with root privileges. It should be noted that in this case, an attacker must be authenticated to access the vulnerable functionality. Unfortunately, that is not always the case. We identified a vulnerability that could allow an unauthenticated, remote attacker to inject arbitrary commands that are executed with root privileges.

So, the following threats can be encountered regarding ZTP:
- ZTP server or/and client spoofing
- Unauthorized access to ZTD service and data in cloud
- Exhaustive Denial of Service on cloud ZTP service
- Privilege escalation on ZTP server
- Eavesdropping
- Insufficient access control in multi-tenant ZTD service

## Cryptography Threats

All known SDN/SD-WAN architecture and reference design guides [6, 7] do not mention a cryptography plane. At the same time cryptographic mechanisms must be employed and employed correctly and secure on all planes: data, control, management, and orchestration (application). Cryptographic mechanisms are essense of SDN/SD-WAN.

In cryptography, it is known two informal rules. The first, the gold rule: "Do not develop your own crypto". And the second one, the silver rule: "Do not implement well-known crypto primitives". Unfortunately, the current state of practical SD-WAN products adds a new rule: "Do not provision your crypto". We found that most of SD-WAN products (fortunately not all) have the following flaws related to cryptographic provisioning:
- Use of hardcoded public-key cryptography key pairs and corresponding certificates that are the same for all customers and can not be replaced
- Use of self-signed certificates for generated public-key cryptography key pairs issued by the SD-WAN product
- Manual installation of self-signed certificates on SD-WAN nodes with no chance to fast revoke them
- Absence of classic CRL and OCSP mechanisms
- Absence of interfaces to be integrated with customer private or public CA

We also were surprised that no explored SD-WAN product uses modern promising cryptographic features, frameworks, mechanisms and tools such as short-lived certificates, Noise, NoiseSocket, Strobe, Disco, WireGuard, SPIFFE. We do not suggest to use them on data plane, but we think that cryptographic primitives are very useful and suitable for cloud-centric applications, northbound, and southbound interfaces.

# 4  CONCLUSIONS

In the paper, we provided and explained most common SD-WAN threats observed through SD-WAN threat modelling and security assessment campaign.

This allowed us to discover many zero-day vulnerabilities in different commercial products. The security assessment was oriented to general security mechanisms and Web interfaces.

The following summarized security issues were identified:
- Silver Peak Unity EdgeConnect [20]
    - Lack of protection against brute-forcing password
    - Web UI leaks software versions
    - Lack of protection against CSRF for REST API
    - Denial of service of Web UI via Slow HTTP attacks
    - REST API leaks software versions
    - Use of default SNMP community strings
    - Access to operating system interface via administrative CLI backdoor
    - Multiple vulnerabilities to reflected XSS
    - Arbitrary file reading via path traversal
- Riverbed SteelConnect [26]
    - Password reset link spoofing via HTTP host header
    - Stored XSS via user name field
    - Denial of service of gateway via slow HTTP attacks
- Cisco (Viptela) SD-WAN
    - OpenSSH leaks system version via warning message
    - Incorrect protection against CSRF for REST API and Web UI
- Viprinet Virtual VPN Hub
    - Stored XSS in CLI via item names
    - TLS server vulnerable to ROBOT attack
- Citrix NetScaler SD-WAN [22]
    - Denial of Service on Web UI via Slow HTTP attacks
    - Multiple stored and reflected XSS
    - Lack of protection against CSRF for REST API and Web UI
    - Absence of function level access control mechanism
    - Multiple command injections
    - Multiple SQL injections
    - Arbitrary file reading via path traversal
    - Unauthorized access to Munin web UI

All these issues were reported to corresponding vendors. Some issues were disclosed via "Full Disclosure Mailing List" [26].

# ACKNOWLEDGMENTS

The authors would like to thank the following members of the "SD-WAN New Hope" team for finding security issues: Maxim Gorbunov, Oleg Broslavsky, Anton Nikolaev, Nikita Oleksov, Nikolay Tkachenko.